\documentclass[aip,pof,reprint,groupedaddress,superscriptaddress]{revtex4-1}
\usepackage[USenglish]{babel}
\usepackage{graphicx}
\usepackage{eqnarray}
\usepackage{amssymb,amsmath,fancyref}
\usepackage{natbib,color}
\usepackage[colorlinks,urlcolor=blue,linkcolor=blue,citecolor=blue]{hyperref}
\usepackage[nooneline]{subfigure}

\def\ppart#1#2{\frac{\partial #1 }{\partial #2}}

\newcommand{\opensquare}{\mbox{$\rlap{$\sqcap$}\sqcup$}}

\newcommand{\dotted}{\protect\mbox{${\mathinner{\cdotp\cdotp\cdotp\cdotp\cdotp\cdotp}}$}}
\newcommand{\dashed}{\protect\mbox{-\;-\;-}}
\newcommand{\full}{{\bf \protect\mbox{---}}}
\newcommand{\dashdot}{{\bf \protect\mbox{--  $\cdot$ --}}}

\begin{document}

\title{Modelling of the subgrid scale wrinkling factor for large-eddy simulation of turbulent premixed combustion}
\author{F. Thiesset}
\affiliation{CNRS ICARE, Avenue de la Recherche Scientifique, 45072 Orl{\'e}ans Cedex 2 France}
\author{G. Maurice}
\affiliation{CNRS ICARE, Avenue de la Recherche Scientifique, 45072 Orl{\'e}ans Cedex 2 France}
\affiliation{University of Orl{\'e}ans, INSA de Bourges, PRISME, EA 4229, 45072 Orl{\'e}eans, France }
\author{F. Halter}
\affiliation{CNRS ICARE, Avenue de la Recherche Scientifique, 45072 Orl{\'e}ans Cedex 2 France}
\affiliation{University of Orl{\'e}ans, INSA de Bourges, PRISME, EA 4229, 45072 Orl{\'e}eans, France }
\author{N. Mazellier}
\affiliation{University of Orl{\'e}ans, INSA de Bourges, PRISME, EA 4229, 45072 Orl{\'e}eans, France }
\author{C. Chauveau}
\affiliation{CNRS ICARE, Avenue de la Recherche Scientifique, 45072 Orl{\'e}ans Cedex 2 France}
\author{I. G{\"o}kalp}
\affiliation{CNRS ICARE, Avenue de la Recherche Scientifique, 45072 Orl{\'e}ans Cedex 2 France}

\begin{abstract}
We propose a model for assessing the unresolved wrinkling factor in LES of turbulent premixed combustion. It relies essentially on a power-law dependence of the wrinkling factor to the filter size and an original expression for the 'active' corrugating strain rate. The latter is written as a product of an efficiency function which accounts for viscous effects and the kinematic constraint of Peters\cite{Peters1986}, by a recent expression for the turbulent strain intensity. Yields functional expressions for the fractal dimension and the inner cut-off length scale, the latter being {\it (i)} filter-size independent and {\it (ii)} consistent with the Damk{\"o}hler asymptotic behaviours at both large and small Karlovitz numbers. A new expression for the wrinkling factor which incorporates finite Reynolds numbers effects is further proposed. Finally, the model is successfully assessed on an experimental filtered database.
\end{abstract}

\maketitle

The key ingredient for modelling the subgrid scale wrinkling factor is the rate of strain which is known to be in part responsible for the corrugation of the flame front\cite{Meneveau1991,Angelberger1998,Colin2000,Charlette2002,Fureby2005,Hawkes2012}. In a recent paper\cite{Thiesset2013}, an expression for the strain intensity $\mathcal{S}$ acting at a scale $r$  was derived (hereafter $r$ denotes either a typical turbulent scale or the LES filter size). In a locally isotropic context, in Kolmogorov units (indicated by an asterisk), the local strain rate reads
\begin{eqnarray}
\mathcal{S}(r^*) = \left[ \frac{1}{r^*} \ppart{}{r^*} \overline{(\Delta {q^*})^2} + \frac{1}{2}\ppart{^2}{{r^*}^{2}} \overline{(\Delta {q^*})^2} \right]^{1/2}. \label{Strain_rate}
\end{eqnarray}
$r^* \equiv r/\eta$ and $\overline{(\Delta {q^*})^2} \equiv \overline{(\Delta {q})^2}/u_K^2 $ where the Kolmogorov scales are $\eta = (\nu^3/\epsilon)^{1/4}$ and $u_K = (\nu \epsilon)^{1/4}$, $\epsilon$ being the mean kinetic energy dissipation rate and $\nu$ the kinematic viscosity. $\overline{(\Delta {q})^2} = \overline{\Delta {u_i}\Delta {u_i}}$ (summation convention applies to double Roman indices) is generally interpreted as the total kinetic energy at a given scale. $\Delta \bullet= \bullet(x+r) - \bullet(x)$ is the spatial increment of the quantity $\bullet$ between two points separated by a distance $r$. The overbar stands for a suitable average. The rate of strain thus appears related to the Laplacian (here expressed in spherical coordinates thanks to local isotropy) of the total kinetic energy at a given scale $\overline{(\Delta {q})^2}$. The transport equation for $\overline{(\Delta {q})^2}$ which follows from an extension of the pioneering work by Refs. \onlinecite{Karman1938,Kolmogorov1941} to slightly inhomogeneous locally isotropic flows writes\cite{Danaila2004a}
\begin{eqnarray}
 \mathcal{I^*} - \frac{3}{4r^*} \overline{\Delta u^* (\Delta {q^*}^2)}  + \frac{3}{2r^*} \ppart{\overline{(\Delta q^*)^2}}{r^*}  = 1.   \label{SBS_q2}
\end{eqnarray}
Eq. \eqref{SBS_q2} describes the dynamical equilibrium between the different ranges of turbulent scales. The first term on LHS of Eq. \eqref{SBS_q2} corresponds to the injection of kinetic energy at large scales through the combined effect of advection, production, turbulent or pressure diffusion. The energy then cascades towards smaller scales in an intermediate range of scales (the inertial range), this process being characterized by the second term on LHS of Eq. \eqref{SBS_q2}. Finally, the last term (hereafter formally written as $\mathcal{V}$) stands for the loss of energy by viscous effects and predominates at the smallest eddies. Remarkable is the fact that the expression for the viscous term appears in the expression for $\mathcal{S}$. Furthermore, at the smallest scales, it is readily shown that $\lim_{r^* \to 0} \mathcal{S}^2(r^*) = \lim_{r^* \to 0} \mathcal{V}(r^*) = 1$. This indicates that at the smallest scales, all the strain is diffused by viscosity and will thus not be efficient enough to corrugate the flame front. In addition to viscous effects, Ref. \onlinecite{Peters1986} suggested that there might be also a kinematic constraint that precludes scales $r$ with characteristic velocity $U_r$ (to be defined later) smaller than the laminar flame speed $S_L$ from wrinkling the flame front. These two key ingredients (viscous + kinematic constraint) naturally lead us to a new definition for the active corrugating strain rate $\mathcal{K}$, viz.
\begin{subequations}
\begin{eqnarray}
\mathcal{K} (r^*)   &=&  \mathcal{C}(r^*) \mathcal{S}(r^*)= \mathcal{C}_1(r^*) \mathcal{C}_2(r^*) \mathcal{S}(r^*) \label{K} \\
\mathcal{C}_1(r^*)  &=& 1-\mathcal{V}(r^*) \label{C1}\\
\mathcal{C}_2(r^*)  &=& \frac{1}{2} \left\lbrace  1+ \textrm{erf} \left[ 3\log_{10} \left( \frac{U_r}{S_L} \right) \right] \right\rbrace \label{C2}
\end{eqnarray}
\end{subequations}
Eq. \eqref{K} characterises the 'active' strain which effectively corrugates the flame front. It is written as the product of an efficiency function by the turbulent strain at a given scale. The efficiency function $\mathcal{C}_1(r^*)$ (Eq. \eqref{C1}) accounts for the rate of strain whose intensity is large enough compared to viscous effects for effectively corrugating the flame front. At large scales, $\mathcal{C}_1(r^*) \to 1$ whereas as $r^* \to 0$, $\mathcal{C}_1(r^*) \to 0$, as expected. On the other hand, the second efficiency function $\mathcal{C}_2(r^*)$ (Eq. \eqref{C2}) whose formulation is largely inspired by that of Ref. \onlinecite{Charlette2002}, is the kinematic constraint which follows from the suggestion of Ref. \onlinecite{Peters1986}. It is worth recalling that unlike previous studies \cite{Meneveau1991,Angelberger1998,Colin2000,Charlette2002} for which $\mathcal{C}$ was assessed by means of canonical flame vortex interactions  DNSs, a plausible phenomenological interpretation of this efficiency function is provided in the present case. 

\begin{figure}
\subfigure[\label{C}]{\includegraphics[scale = 0.34]{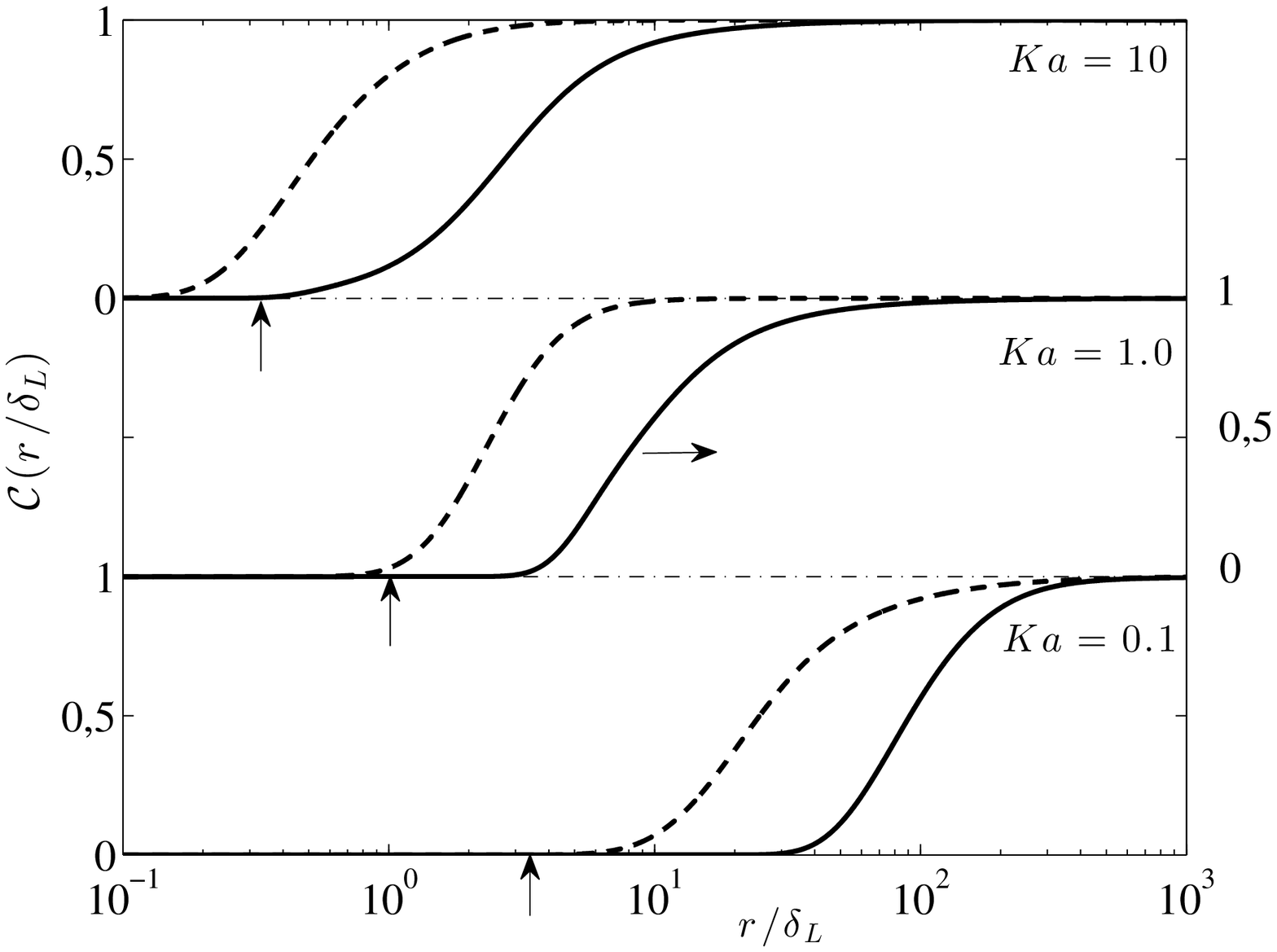}}
\subfigure[\label{Strain}]{\includegraphics[scale = 0.34]{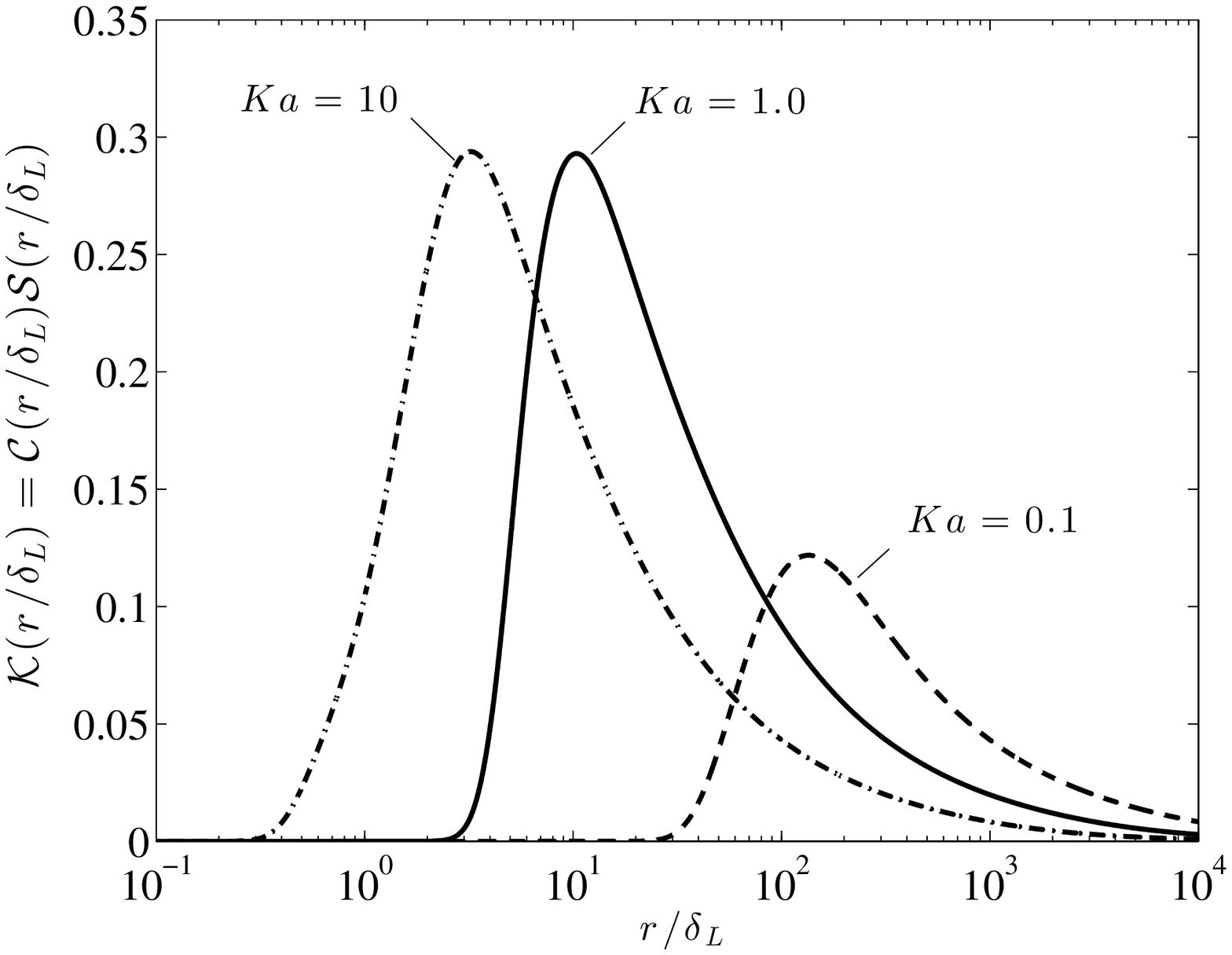}}
\subfigure[\label{eta_i}]{\includegraphics[scale=.34]{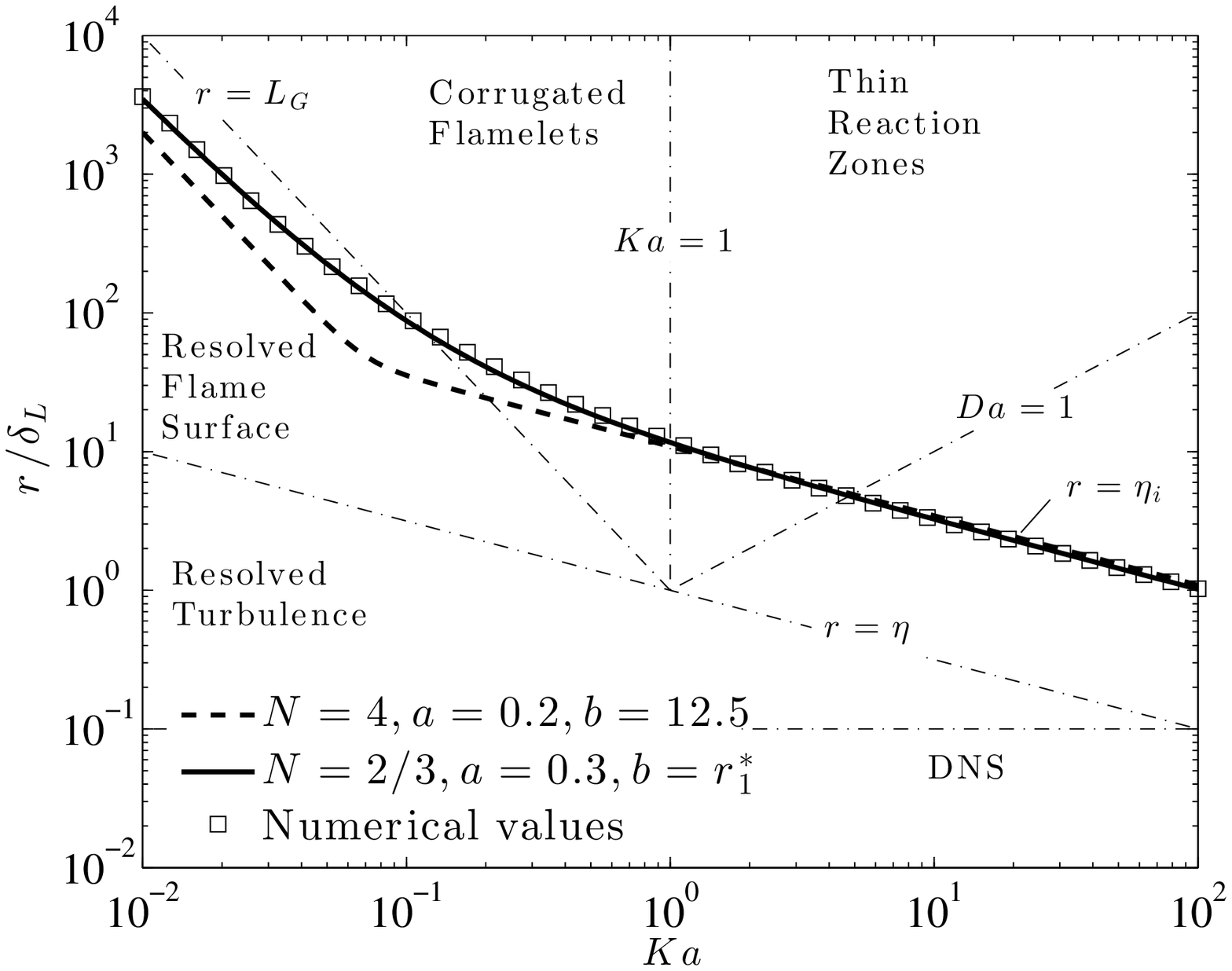}}
\subfigure[\label{ST_real}]{\includegraphics[scale=.34]{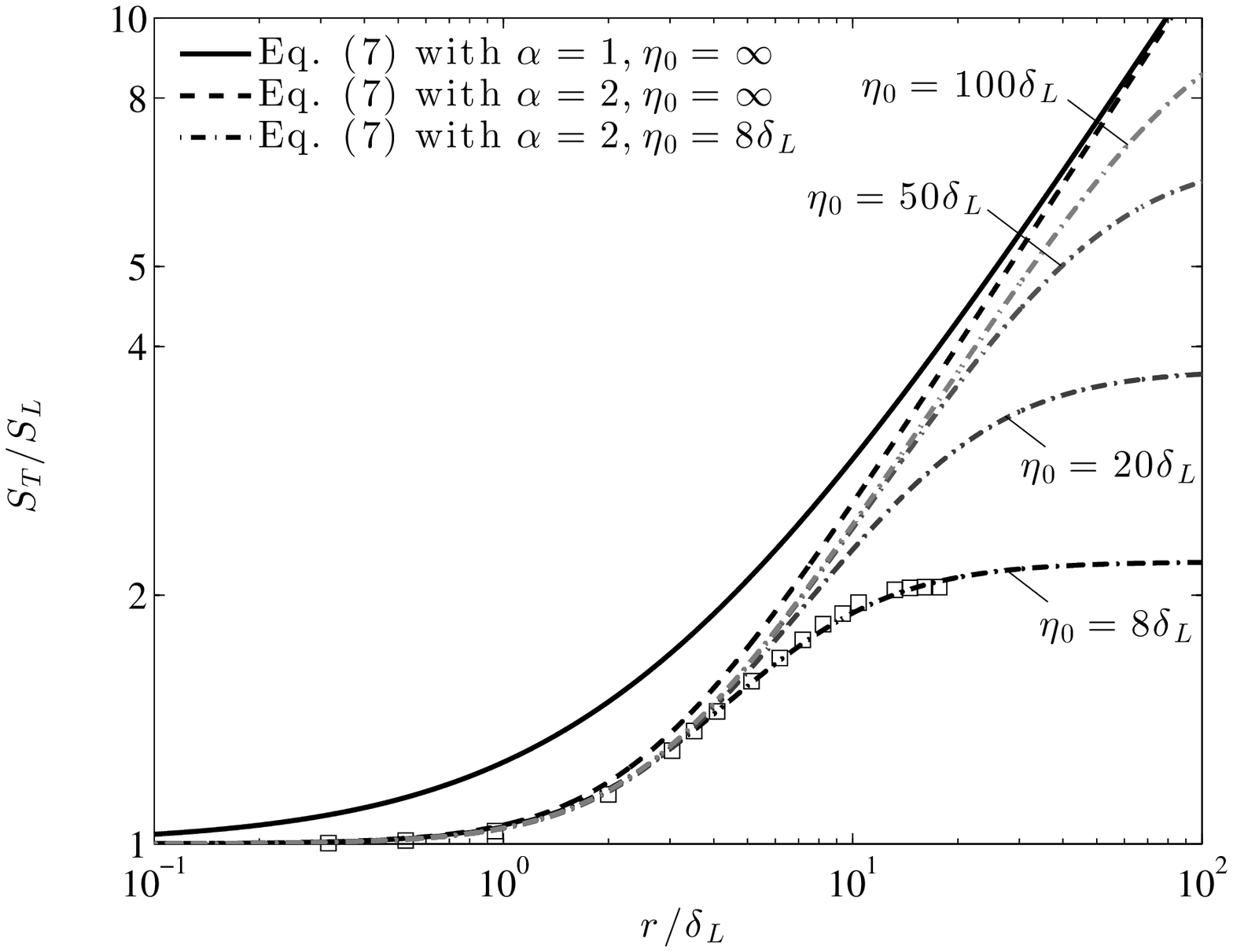}}
\caption{(a) Comparison of the proposed efficiency function ($\full$) $\mathcal{C}(r/\delta_L)$  as a function of filter size $r/\delta_L$ with that of Ref. \onlinecite{Charlette2002} ($\dashed$) . (b) Effective rate of strain $\mathcal{C}(r/\delta_L)\mathcal{S}(r/\delta_L)$ for three different Karlovitz numbers,  $\dashed$ $Ka = 0.1$; $\full$ $Ka = 1.0$; $\dashdot$ $Ka = 10$. (c) Inner cut-off length-scale $\eta_i$ in the LES regime diagram\cite{Pitsch2002}. $\opensquare$ corresponds to the present numerical values computed from $\max[\mathcal{K(r^*)}]$. Also displayed is Eq. \eqref{Hawkes} with ($\dashed$) $N = 4$, $a = 0.2$ and $b = 12.5$ (Ref. \onlinecite{Hawkes2012}), ($\full$) $N = \frac{2}{3}$, $a =0.3$ and $b = r_1^*$ (present study). (d) $S_T/S_L$ as a function of filter size $r/\delta_L$. $\opensquare$ DNS data of Ref. \onlinecite{Hawkes2012} which are compared to that modelled by Eq. \eqref{ST_SL} with $\eta_i$ and $\beta$ predicted using $N = \frac{2}{3}$, $a = 0.3$ and $b = r_1^*$. $\full$ $\alpha = 1, \eta_o = \infty$, $\dashed$ $\alpha = 2, \eta_o = \infty$, $\dashdot$ $\alpha = 2, \eta_o/\delta_L = 8$. For all curves, the Karlovitz number was set to $Ka = 11$. Also displayed in (d) is Eq. \eqref{ST_SL} with $\eta_o/\delta_L = 20, 50$ and $100$.}
\end{figure}

At this stage, a rather realistic functional for $\overline{(\Delta {q^*})^2}$ and $U_r$ needs to be further employed for an analytical expression for $\mathcal{K} (r^*)$ to be derived. In previous efforts\cite{Peters1986,Poinsot1991a,Meneveau1991,Angelberger1998,Colin2000,Charlette2002,Hawkes2012}, inertial range relations were used for describing  $\overline{(\Delta {q^*})^2}$ and $U_r$, i.e.  $\overline{(\Delta {q^*})^2} \propto U_r^2 \propto r^{2/3}$. Clearly, such an hypothesis might not be applicable since the Reynolds numbers generally encountered in practical situations is not sufficiently large for the inertial range to be discernible. In order to provide a more appropriate expression for both $\overline{(\Delta {q^*})^2}$ and $U_r$ which accounts for notably finite Reynolds number effects, we first recall that in the dissipative range, under the constraint of local isotropy, we have $\overline{(\Delta {q^*})^2} = {r^*}^2/3$. In the inertial range, the Kolmorogov's relation reads $\overline{(\Delta {q^*})^2} = C_q {r^*}^{2/3}$ ($C_q$ is related to the Kolmogorov constant $C_u$ by $C_q = 11C_u/3$ and will hereafter be set to $22/3$\cite{Antonia2003}). Finally, at large scales $\overline{(\Delta {q^*})^2} = 2\overline{ {q^*}^2}$. $\overline{ {q^*}^2}$ is related to the turbulent Reynolds number $Re_t = u'L_t/\nu$ ($L_t$ is the integral length-scale and $u'$ a typical velocity fluctuation) through the relation $\overline{ {q^*}^2} = 3Re_t^{1/2}$. Hereafter, a value of 300 for $\overline{ {q^*}^2}$ is prescribed as an illustration. Following an elegant interpolation first proposed by Batchelor\cite{Batchelor1951}, these asymptotic scalings can be matched together in a parametric equation of the form \cite{Antonia2003}
\begin{eqnarray}
\overline{(\Delta {q^*})^2}  =  \frac{{r^*}^2}{3} \left[1 + \left(\frac{r^*}{r_1^*} \right)^2 \right]^{-2/3} \left[1 + \left(\frac{r^*}{r_2^*} \right)^2 \right]^{-1/3} \label{s2q}
\end{eqnarray}
where $r_1^* = (3 C_q)^{3/4}$ is the cross over between the viscous and inertial range, whilst the cross-over between large and inertial scales is given by $r_2^* = (2\overline{{q^*}^2}/C_q)^{3/2}$. Even though this parametric expression is built using asymptotic relations, it appears to be well suited for describing $\overline{(\Delta {q^*})^2}$ even at low Reynolds numbers\cite{Antonia2003}. Then, for $U_r$, one can simply write
\begin{eqnarray}
\frac{U_r}{S_L}= \left[\frac{\overline{(\Delta {q^*})^2}}{6}\right]^{1/2} Ka^{1/2}. \label{Ur}
\end{eqnarray}
Appears the Karlovitz number $Ka = (u_K/S_L)^2 = (\delta_L/\eta)^2$ where $\delta_L = D/ S_L$ is the laminar flame thickness, with $D$ the fresh gas diffusivity. It is worth noting that this expression for $U_r$ might be preferably used to construct spectral diagram following the lines of Ref. \onlinecite{Poinsot1991a}. By further using Eq. \eqref{s2q} yields analytical expression for $\mathcal{V}$ or $\mathcal{S}$, and consequently $\mathcal{K}$. In Fig. \ref{C}, the proposed efficiency function $\mathcal{C}(r) = \mathcal{C}_1(r)\mathcal{C}_2(r)$ is compared to that of Ref. \onlinecite{Charlette2002} 
\begin{eqnarray}
\mathcal{C}_{ch} = \frac{1}{4} \left\lbrace  1+ \textrm{erf} \left[ 0.6\log \left( \frac{r}{\delta_L}\right)  - \left(\frac{U_r}{S_L}\right)^{-1/2}\right] \right\rbrace  \left\lbrace  1+ \textrm{erf} \left[ 3\log_{10} \left( 2\frac{U_r}{S_L} \right) \right] \right\rbrace
\end{eqnarray}

which is an extension of the DNS based expression provided by Ref. \onlinecite{Colin2000} to account for the kinematic constraint discussed previously. To plot $\mathcal{C}_{ch}$, use was made of Eq. \eqref{Ur} for $U_r$. As the Karlovitz number increases, both expressions progressively drift towards smaller $r/\delta_L$, as expected from the decreasing ratio between $\delta_L$ and the Kolmogorov length-scale $\eta$. Note that at low Karlovitz number, the efficiency function tends to zero at scales larger than $\eta$ (denoted by vertical arrows) revealing that the kinematic constraint $\mathcal{C}_2$ predominates by comparison with $\mathcal{C}_1$. At high Karlovitz numbers, the contrary is observed indicating that viscous effects are mostly perceptible. Departures between $\mathcal{C}$ and $\mathcal{C}_{ch}$ are discernible, irrespectively of the Karlovitz number. A careful analysis at a Karlovitz number of 10 indicates that $\mathcal{C}_{ch}$ is about 20\% at $r = \eta$, which seems rather non-physical since turbulence statistically ceases to be active for scales smaller than $\eta$. This is likely due to the extension of the results emanating from canonical vortex flame interaction studies to a fully turbulent flow. Moreover, when  searching a plausible parametric expression for $\mathcal{C}$, Refs. \onlinecite{Meneveau1991,Colin2000,Charlette2002} implicitly supposes that $\delta_L$ was the relevant length-scale, whereas viscous effects are more likely to scale with the Kolmogorov length-scale $\eta$. 

The active strain rate for Karlovitz number of $0.1$, $1.0$ and $10$ is displayed in Fig. \ref{Strain}. At low Karlovitz number ($\eta < \delta_L$), the strain is active at rather large-scales whilst it drifts towards smaller scales as $Ka$ increases. The maximum value of $\mathcal{K}$ is significantly reduced at low Karlovitz number as a consequence of the kinematic constraint which does not allow scales with characteristic velocity smaller than $S_L$ to exist. Remarkable is the maximum magnitude of the Kolmogorov normalized rate of strain at high Karlovitz numbers which is about $0.29$, this value being extremely close to the RMS of the strain acting on a material line found by Ref. \onlinecite{Yeung1990} from DNSs. 

The idea in deriving an expression for $\mathcal{K}$ is to provide an estimation for the inner cut-off length-scale $\eta_i$, i.e. the smallest characteristic length scale of the flame front wrinkling. Indeed, following a (mono) fractal approach\cite{Meneveau1991,Colin2000,Charlette2002,Fureby2005,Hawkes2012} , the subgrid scale wrinkling factor $\Xi$ (the ratio of the total to the resolved flame surface density), or equivalently the normalized subgrid scale reactants consumption speed $S_T/S_L$ is generally related to $\eta_i$ by
\begin{eqnarray}
\Xi = \frac{S_T}{S_L} = \left[1+\left(\frac{r}{\eta_i}\right)^\alpha\right]^{\beta/\alpha}  \left[1+\left(\frac{r}{\eta_o}\right)^\alpha\right]^{-\beta/\alpha} \label{ST_SL}
\end{eqnarray}
where $\beta = D_f - 2 $ with $D_f$ the fractal dimension. The exponent $\alpha$ is introduced here for the sake of generality and is likely to be related to the exponent $\alpha$ introduced by Ref. \onlinecite{Pocheau1994} in the context of a scale invariance analysis of a propagating flame front.  The outer length-scale $\eta_o$ which is {\it a priori} proportional to the integral turbulent scale is also introduced to account for finite Reynolds number effects. Whilst previous models\cite{Colin2000,Charlette2002,Fureby2005,Hawkes2012} generally use a value of 1 for $\alpha$ and supposes $\eta_o = \infty$, no particular prediction can be drawn at this stage. This point will be examined later when results are compared to DNS data. 

To further proceed, we now have to relate the inner cut-off length-scale $\eta_i$ to the rate of strain at a given scale. Generally\cite{Angelberger1998,Colin2000,Charlette2002,Fureby2005,Hawkes2012}, $\eta_i$  is inferred from a dynamical equilibrium hypothesis between production and destruction of subgrid scale flame surface density. Here, we propose an alternative approach, conjecturing that $\eta_i$ corresponds to the scale $r$ at which the effective turbulent strain rate $\mathcal{K}(r)$ is maximum, i.e. $\eta_i \equiv r$ such as $\mathcal{K}(r) = \max\left[\mathcal{K}(r)\right]$. We may justify this choice by arguing that the scale $\eta_i$ at which $\mathcal{K}$ is maximum corresponds to the scale at which the strain characteristic time scale is the smallest by comparison with the viscous (i.e. the Kolmogorov time scale) characteristic time scale. The maximum value of $\mathcal{K}$ at high Karlovitz number which is consistent with the estimation of Ref. \onlinecite{Yeung1990} also encourages us in adopting this definition. Furthermore, a practical advantage of using this definition is that the inner-cut off length-scale then remains filter size independent (in agreement with the DNS result of Ref. \onlinecite{Hawkes2012}, see Fig. 11) and it is characteristic of a physical rather than a numerical quantity, following the suggestion of Ref. \onlinecite{Hawkes2012}. Since an exact solution for the scale of maximum effective strain is hardly derivable, we prefer assess $\eta_i $ numerically by searching the zero crossing of $\partial \mathcal{K}/\partial{r}$ and then find an appropriate functional describing its evolution as a function of pertinent flame and/or turbulence parameters.

Results are presented in Fig. \ref{eta_i}, where numerical values for $\eta_i/\delta_L$ are plotted as a function of the Karlovitz number. Noticeable is the transition between two different regimes at low and high Karlovitz number which appears at $Ka \approx 0.1$. At low Karlovitz numbers, $\eta_i/\delta_L \propto Ka^{-2}$, meaning that $\eta_i$ is proportional to the Gibson length-scale\cite{Peters1986} $L_G$ . On the other hand, it appears that at high Karlovitz, $\eta_i/\delta_L = r_1^* Ka^{-1/2}$, i.e. is equal to the cross-over length-scale between the viscous and inertial range introduced previously (Eq. \eqref{s2q}). It is worth recalling that the latter expression for $\eta_i$ has already been proposed by Ref. \onlinecite{Kobayashi2002}. There is thus a transition between two regimes, the first one at low Karlovitz numbers where the kinematic constraint $\mathcal{C}_2$ dominates (as noted in Fig. \ref{C}) and the inner cut-off scales with the Gibson length-scale. For this range of Karlovitz numbers, the flame front is a highly active scalar, whose propagation speed acts as a filter precluding fresh gas pockets with characteristic scales smaller than $L_G$ to exist. The second regime, at large Karlovitz number, indicates that the cut-off scale is proportional to $\eta$ (or similarly the Obukhov-Corrsin length-scale notwithstanding the constancy of the Schmidt number). In this regime, the flame front thus behaves rather like a passive scalar\citep{Hawkes2012}. These two different regimes corresponds respectively to the Damkh{\"o}ler large-scale and small-scale asymptotic limits as discussed by Ref. \onlinecite{Pitsch2002} and further recovered analytically by Ref. \onlinecite{Hawkes2012} on the basis of both dimensional and dynamical arguments. Ref. \onlinecite{Hawkes2012} further proposed the following functional to "smoothly" interpolate these two regimes in a single expression, viz.
\begin{eqnarray}
\frac{\eta_i}{\delta_L} = \left[ \left(aKa^{-2}\right)^N+\left(bKa^{-1/2}\right)^N\right]^{1/N}. \label{Hawkes}
\end{eqnarray}
The magnitude of $N$ characterizes the sharpness of the transition (a value of 4 was chosen by Ref. \onlinecite{Hawkes2012}), and $a = 0.2$, $b = 5.5Sc^{-3/4} \approx 12.5$ (providing a Schmidt number $Sc = \nu/D$ of 0.335 for the hydrogen-air mixture at an equivalence ratio of 0.7 and a temperature of 700K as per Ref. \onlinecite{Hawkes2012}) were set {\it ad hoc} by Ref. \onlinecite{Hawkes2012}. When using original values for $a,b$ and $N$, the functional of Ref. \onlinecite{Hawkes2012} appears to differ significantly from the numerical values especially at low Karlovitz numbers (see Fig. \ref{eta_i}). However, using present values for $a=0.3$ and $b = r_1^*$ yields $N=2/3$ for Eq. \eqref{Hawkes} to fit almost perfectly the numerical assessment of $\eta_i$. Therefore, our approach allows to predict the model constants $N$, $a$ and $b$ of Ref. \onlinecite{Hawkes2012} on some physical basis. Unlike Ref. \onlinecite{Fureby2005} for which an empirical expression for $\beta$ was employed, Ref. \onlinecite{Hawkes2012} demonstrated that the fractal dimension $D_f$ should transit from a value of 7/3 at low Karlovitz number corresponding to the fractal dimension of a turbulent/non turbulent interface, to a value of 8/3 at high Karlovitz numbers, the latter value being generally observed for passive scalar fields in fully turbulent flows. To characterize this transition, Ref. \onlinecite{Hawkes2012} proposed the following parametric relation
\begin{eqnarray}
\beta = D_f - 2 = \frac{1}{3} + \frac{1}{3} \frac{\left(bKa^{-1/2}\right)^N}{\left(aKa^{-2}\right)^N + \left(bKa^{-1/2}\right)^N }. \label{beta_fit}
\end{eqnarray}
Plugging the prediction for $\eta_i$ and $\beta$ as given by Eqs. \eqref{Hawkes} and \eqref{beta_fit} into Eq. \eqref{ST_SL} yields an estimation of the sub-grid scale wrinkling factor $\Xi$ or identically $S_T/S_L$ as a function of filter size $r$ and Karlovitz number. Results are presented in Fig. \ref{ST_real}. When compared to the DNS results of Ref. \onlinecite{Hawkes2012}  (see Fig. \ref{ST_real}), it is observed that, keeping $\eta_i$ and $\beta$ unchanged, a value of 2 for $\alpha$ and $\eta_o/\delta_L = 8$ (i.e. about three times larger than the integral length-scale) are much more suitable. Speculatively, the fact that a value of $2$ for $\alpha$ appears more appropriate suggests that the fractal facet of turbulent flames is most likely related to its surface (i.e.  $r^2$) rather than to its scale ($r^1$). At this stage, we have to emphasize that the parameters $\beta = D_f -2$ and $\eta_i$ are kept constant to obtain the curves in Fig. \ref{ST_real}, whereas the apparent slope in the 'inertial' range of filter size appears less steeper in the DNS. This indicates that in most of practical situations, the scale separation between $\eta_o$ and $\eta_i$ (or equivalently the turbulent Reynolds number) is not sufficiently large for a proper fractal dimension to be unambiguously inferred and its estimation is clearly biased by some so-called finite Reynolds number effects. Consequently, finite Reynolds number effects are likely to shed doubts on most of the experimental or numerical estimations of $\beta$ and $\eta_i$ and the quest for an universal value or universal evolution is worth being revisited.

To further validate the reliability of the present model, {\it a priori} tests are provided by comparing to experiments. The experimental set-up has been fully detailed in Refs. \onlinecite{Fragner2014,Fragner2014a} and is briefly recalled here. A methane-air premixed Bunsen flame at a pressure of 0.3MPa and equivalence ratio of 0.6 is considered for this test. High-speed Mie-scattering tomography on organic oil droplets, allows the instantaneous flame front to be tracked by classical contour edge detection. Yields the progress variable $c$, which is by definition equal to 0 and 1 in the unburned and burned gas respectively. Intense turbulence is generated by a multi-scale grid and has been fully characterized by means of hot-wire measurements by Ref. \onlinecite{Fragner2014a}. By definition, the Reynolds average of the Flame Surface Density (FSD) is $\overline{|\nabla c|}$, whilst the resolved surface density is $\overline{|\nabla \langle c \rangle|}$, where $c$ is the progress variable and the brackets stand for filtered quantities using a (Reynolds) gaussian filter. Then, the total FSD is reconstructed by multiplying $\overline{\nabla \langle c \rangle}$ by the wrinkling factor $\Xi$ as given by Eq. \eqref{ST_SL}. For the present case, $Ka$ is decreasing from a value of 6.3 to 1.5 and the integral length-scale $L_t$ varies between 4.2mm to 6.3mm as the streamwise distance from the burner $x$ increases. 

\begin{figure}
\begin{center}
\subfigure[\label{Model_40Hf}]{\includegraphics[scale=.34]{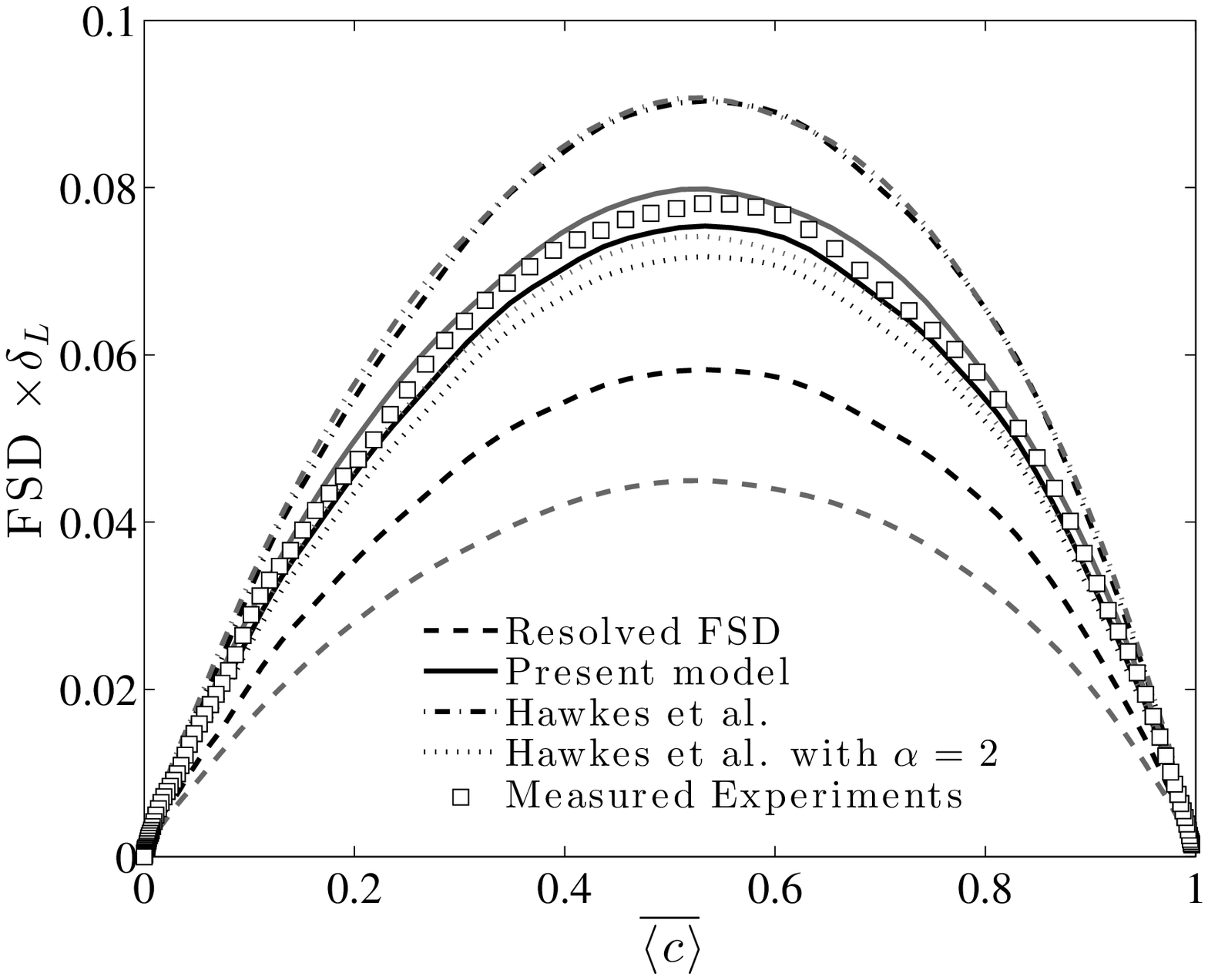}}
\subfigure[\label{Model_70Hf}]{\includegraphics[scale=.34]{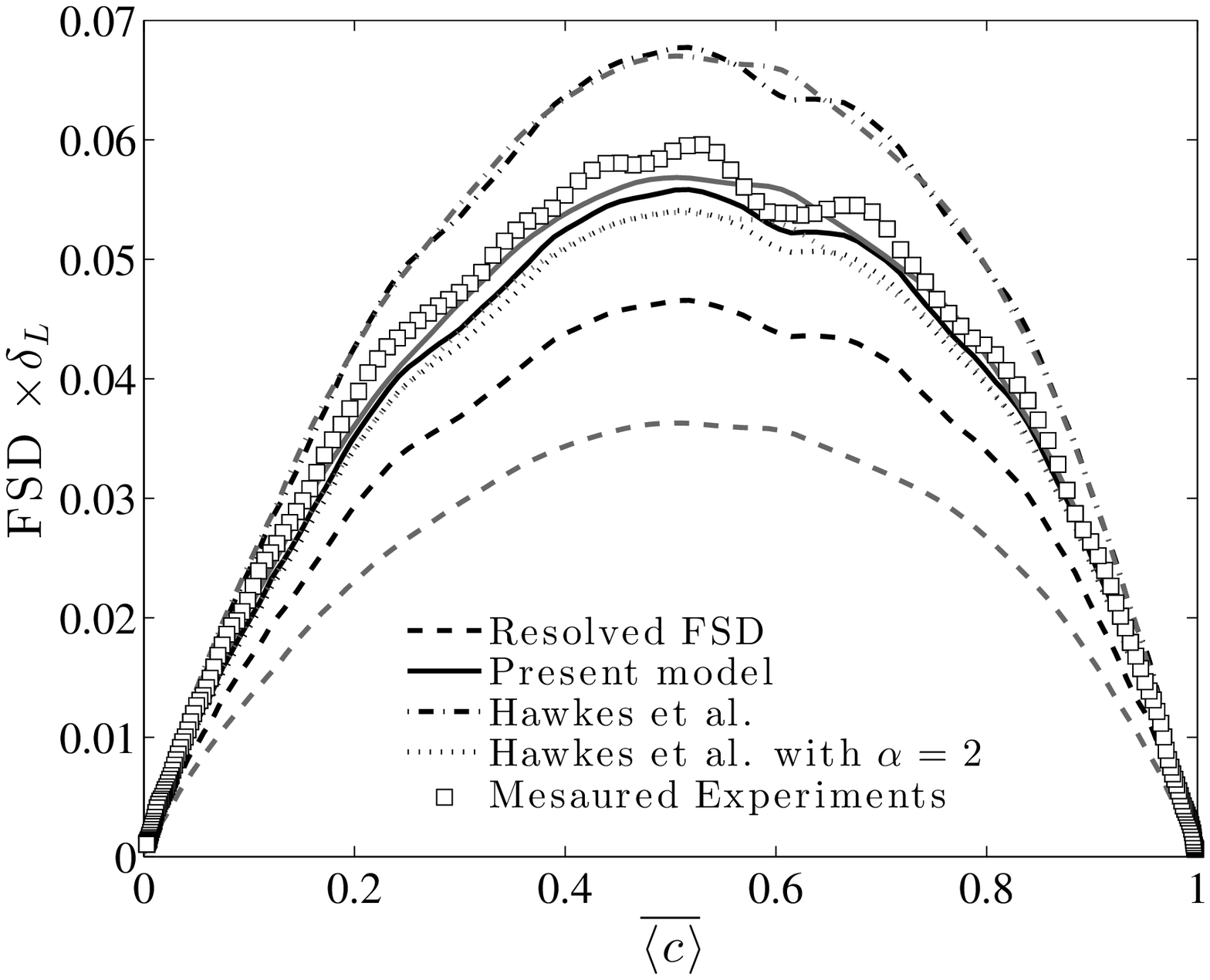}}
\end{center}
\caption{FSD as a function of the mean filtered progress variable $\overline{\langle c \rangle}$, for two distinct streamwise locations in the flow, (a) $x/H_f = 0.4$ and (b) $x/H_f = 0.7$ (the flame height $H_f$ is defined as the location $x$ where $\overline{c} = 5\%$). $\opensquare$ measured FSD, $\dashed$ resolved FSD, $\full$ present model $N = 2/3$, $a = 0.3$, $b = r_1^*$, $\eta_0 = 3L_t$ and $\alpha =2$, $\dashdot$ Hawkes et al.\cite{Hawkes2012} model $N = 4$, $a = 0.2$, $b = 12.5$, $\eta_0 = \infty$ and $\alpha = 1$, $\dotted$ Hawkes et al.\cite{Hawkes2012} model with $N = 4$, $a = 0.2$, $b = 12.5$, $\eta_0 = \infty$ and $\alpha = 2$. The black curves corresponds to a filter size of $8dx$, whilst gray curves to a filter size of $16dx$.}
\end{figure}

Results are presented in Figs. \ref{Model_40Hf} and \ref{Model_70Hf} where the measured FSD is first compared to that resolved using a filter size of $8$ and $16dx$. Noticeable is the magnitude of the resolved FSD which is attenuated by about 25\% and 50\%, when the filter size increases from $8dx$ to $16dx$ ($dx = 0.105$mm$\approx 0.8\delta_L$ is the spatial resolution of the laser tomography). The present model yields very encouraging results since the reconstructed FSD agrees almost perfectly with that inferred from experiments, irrespectively of the filter size and streamwise distance (Figs. \ref{Model_40Hf} and \ref{Model_70Hf}). However, when Hawkes et al.\cite{Hawkes2012} values for the constants $N$, $a$, $b$, $\eta_0$ and $\alpha$ are prescribed, one notes slightly overestimated values by about 10-15\% depending on the location in the flow. This departure is largely attributed to the chosen value for $\alpha$. Indeed, with $\alpha = 2$, keeping the other constants in the Hawkes et al.\cite{Hawkes2012} model, yields a correct estimation of the total FSD. Here again, this suggests that a value of 2 for $\alpha$ is more appropriate. Although results are not presented here, the models have been tested for other Karlovitz and Reynolds numbers, leading to similar deductions. 

In summary, six distinct outcomes emerge from the present study. 

{\it (i) } An analytical expression for the efficiency function is proposed on the basis of some physical reasoning arguments. It accounts for viscous effects which dominates at high Karlovitz number as well as a kinematic constraint {\`a} la Peters\cite{Peters1986} whose effect is dominant at low Karlovitz numbers. These two distinct regimes correspond respectively to the Damkh{\"o}ler small and large-scale asymptotic limits. In the small-scale asymptotic limit, the flame front behaves as a passive scalar and the maximum effective strain predicted by the present model is in perfect agreement with the value of 0.28 inferred by Ref. \onlinecite{Yeung1990}. 

{\it (ii) } The inner cut-off length-scale follows from the conjecture that $\eta_i$ is the scale at which the active strain rate is maximum. As a consequence of the definition of $\mathcal{C}$, $\eta_i$ also reveals two different scaling with Karlovitz number corresponding to the two aforementioned regimes. It is then observed that at low Karlovitz number, the cut-off corresponds to the Gibson length-scale, as suggested by Peters\cite{Peters1986}, whilst at high Karlovitz numbers, the cut-off is the cross-over length-scale between viscous and inertial ranges (i.e. $r_1^* = (3C_q)^{3/4}$) in agreement with Ref. \onlinecite{Kobayashi2002}. 

{\it (iii) }  The present approach allows to estimate the constants $a,b,N$ in the model of Ref. \onlinecite{Hawkes2012} on the basis of some physical arguments. It is thus proven that $a=0.3$, $b = r_1^*$ and $N = 2/3$ are more suitable. 

{\it (iv) }  A new expression for the wrinkling factor (Eq. \eqref{ST_SL}) is introduced, revealing additional parameters $\eta_0$ and $\alpha$, the latter being inspired by the scale invariance analysis of Ref. \onlinecite{Pocheau1994}, the former related to the integral length scale to account for finite Reynolds number effects. This expression compares favourably well with the DNS results of Ref. \onlinecite{Hawkes2012} when a value of $2$ is chosen for $\alpha$ which highlights that the fractal nature of turbulent flames is likely to be related to its surface rather than to its scale. 

{\it (v) }  A careful analysis of Fig. \ref{ST_real} shows that in most of practical situations, the Reynolds number is not sufficiently large for a reliable value of the fractal dimension to be inferred. At finite Reynolds numbers, the estimation is biased, and the apparent fractal dimension systematically appears less steeper than the asymptotic value.

{\it (vi) }  {\it A priori} tests are provided by comparing modelled FSDs to that measured in lean methane-air Bunsen flames. These tests give strong support in favour of the present model.

\vspace{4pt}
The financial support from the Agence National de la Recherche under the project IDYLLE is gratefully acknowledged. We are also thankful to the CNRS, the University of Orl{\'e}ans, and the French Government Program "Investissements d'avenir" through the LABEX CAPRYSSES.
\bibliography{NC15}

\end{document}